\documentclass[twoside,twocolumn,9pt]{article}
\usepackage{extsizes}
\usepackage[super,sort&compress,comma]{natbib} 
\usepackage[version=3]{mhchem}
\usepackage[left=1.5cm, right=1.5cm, top=1.785cm, bottom=2.0cm]{geometry}
\usepackage{balance}
\usepackage{times,mathptmx}
\usepackage{sectsty}
\usepackage{graphicx} 
\usepackage{lastpage}
\usepackage[format=plain,justification=justified,singlelinecheck=false,font={stretch=1.125,small,sf},labelfont=bf,labelsep=space]{caption}
\usepackage{float}
\usepackage{fancyhdr}
\usepackage{fnpos}
\usepackage[english]{babel}
\addto{\captionsenglish}{%
  
}
\usepackage{array}
\usepackage{droidsans}
\usepackage{charter}
\usepackage[T1]{fontenc}
\usepackage[usenames,dvipsnames]{xcolor}
\usepackage{setspace}
\usepackage[compact]{titlesec}
\usepackage{hyperref}
\usepackage{bm}
\usepackage[normalem]{ulem}
\usepackage{color}
\usepackage{xcolor} 

\usepackage{epstopdf}

\definecolor{cream}{RGB}{222,217,201}

\begin{document}

\pagestyle{fancy}
\thispagestyle{plain}
\fancypagestyle{plain}{

\fancyhead[C]{\includegraphics[width=18.5cm]{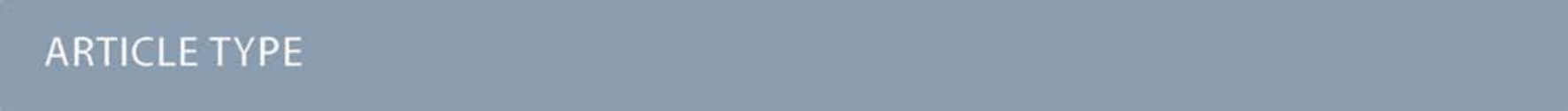}}
\fancyhead[L]{\hspace{0cm}\vspace{1.5cm}\includegraphics[height=30pt]{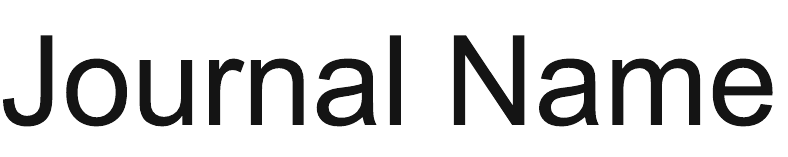}}
\fancyhead[R]{\hspace{0cm}\vspace{1.7cm}\includegraphics[height=55pt]{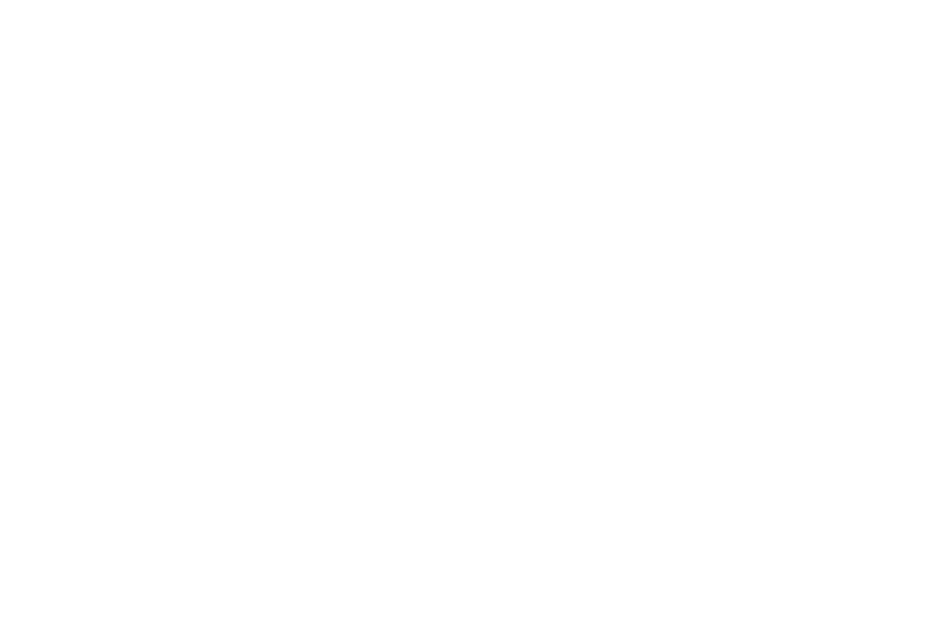}}
\renewcommand{\headrulewidth}{0pt}
}

\makeFNbottom
\makeatletter
\renewcommand\LARGE{\@setfontsize\LARGE{15pt}{17}}
\renewcommand\Large{\@setfontsize\Large{12pt}{14}}
\renewcommand\large{\@setfontsize\large{10pt}{12}}
\renewcommand\footnotesize{\@setfontsize\footnotesize{7pt}{10}}
\renewcommand\scriptsize{\@setfontsize\scriptsize{7pt}{7}}
\makeatother

\renewcommand{\thefootnote}{\fnsymbol{footnote}}
\renewcommand\footnoterule{\vspace*{1pt}%
\color{cream}\hrule width 3.5in height 0.4pt \color{black} \vspace*{5pt}} 
\setcounter{secnumdepth}{5}

\makeatletter 
\renewcommand\@biblabel[1]{#1}            
\renewcommand\@makefntext[1]%
{\noindent\makebox[0pt][r]{\@thefnmark\,}#1}
\makeatother 
\renewcommand{\figurename}{\small{Fig.}~}
\sectionfont{\sffamily\Large}
\subsectionfont{\normalsize}
\subsubsectionfont{\bf}
\setstretch{1.125} 
\setlength{\skip\footins}{0.8cm}
\setlength{\footnotesep}{0.25cm}
\setlength{\jot}{10pt}
\titlespacing*{\section}{0pt}{4pt}{4pt}
\titlespacing*{\subsection}{0pt}{15pt}{1pt}

\fancyfoot{}
\fancyfoot[LO,RE]{\vspace{-7.1pt}\includegraphics[height=9pt]{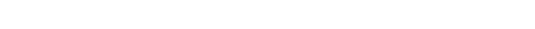}}
\fancyfoot[CO]{\vspace{-7.1pt}\hspace{13.2cm}\includegraphics{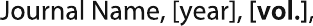}}
\fancyfoot[CE]{\vspace{-7.2pt}\hspace{-14.2cm}\includegraphics{head_foot/RF}}
\fancyfoot[RO]{\footnotesize{\sffamily{1--\pageref{LastPage} ~\textbar  \hspace{2pt}\thepage}}}
\fancyfoot[LE]{\footnotesize{\sffamily{\thepage~\textbar\hspace{3.45cm} 1--\pageref{LastPage}}}}
\fancyhead{}
\renewcommand{\headrulewidth}{0pt} 
\renewcommand{\footrulewidth}{0pt}
\setlength{\arrayrulewidth}{1pt}
\setlength{\columnsep}{6.5mm}
\setlength\bibsep{1pt}

\makeatletter 
\newlength{\figrulesep} 
\setlength{\figrulesep}{0.5\textfloatsep} 

\newcommand{\topfigrule}{\vspace*{-1pt}%
\noindent{\color{cream}\rule[-\figrulesep]{\columnwidth}{1.5pt}} }

\newcommand{\botfigrule}{\vspace*{-2pt}%
\noindent{\color{cream}\rule[\figrulesep]{\columnwidth}{1.5pt}} }

\newcommand{\dblfigrule}{\vspace*{-1pt}%
\noindent{\color{cream}\rule[-\figrulesep]{\textwidth}{1.5pt}} }

\makeatother

\twocolumn[
  \begin{@twocolumnfalse}
\vspace{3cm}
\sffamily
\begin{tabular}{m{4.5cm} p{13.5cm} }

\includegraphics{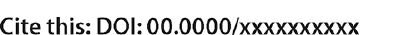} & \noindent\LARGE{\textbf{Direct Measurement of Lighthill's Energetic Efficiency of a Minimal Magnetic Microswimmer$^\dag$}} \\
 & \vspace{0.3cm} \\

& \noindent\large{Carles Calero,$^{\ddag}$\textit{$^{a,b}$} Jos\'e Garc\'ia-Torres,$^{\ddag}$\textit{$^{a,b}$}, Antonio Ortiz-Ambriz,\textit{$^{a,b}$} Francesc Sagu\'{e}s,\textit{$^{c,b}$}, Ignacio Pagonabarraga,\textit{$^{a,d,e}$} and Pietro Tierno
\textit{$^{a,b,e *}$}} \\
\includegraphics{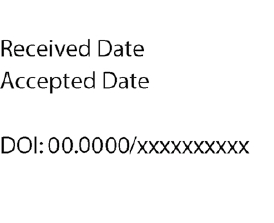} & \\

\end{tabular}

 \end{@twocolumnfalse} \vspace{0.6cm}

  ]

\renewcommand*\rmdefault{bch}\normalfont\upshape
\rmfamily
\section*{}
\vspace{-1cm}


\footnotetext{\textit{$^{a}$~Departament de F\'isica de la Mat\`eria Condensada, Universitat de Barcelona, 08028 Barcelona, Spain.}}
\footnotetext{\textit{$^{b}$~Institut de Nanoci\`encia i Nanotecnologia,
Universitat de Barcelona, 08028 Barcelona , Spain. }}
\footnotetext{\textit{$^{c}$~Departament de Ci\`{e}ncia de Materials i Qu\'{i}mica F\'{i}sica, Universitat de Barcelona, Barcelona, Spain.}}
\footnotetext{\textit{$^{d}$~CECAM, Centre Europ\'een de Calcul Atomique et Mol\'eculaire, \'Ecole Polytechnique F\'ed\'erale de Lasuanne, Batochime, Avenue Forel 2, Lausanne, Switzerland. }}
\footnotetext{\textit{$^{e}$~Universitat de Barcelona, Institute of Complex Systems (UBICS), Universitat de Barcelona, 08028 Barcelona, Spain.}}
\footnotetext{* E-mail: ptierno@ub.edu}

\footnotetext{\dag~Electronic Supplementary Information (ESI) available: Two experimental videos (.WMF)
showing the 
dynamics of the nanorod-colloid micropropeller
played at different speed. See DOI: 00.0000/00000000.}

\footnotetext{\ddag These authors contributed equally to this work.}



\sffamily{\textbf{The realization of artificial microscopic swimmers able to propel in viscous fluids is an emergent research field of fundamental interest and vast technological applications. For certain functionalities, the efficiency of the microswimmer in converting the input power provided through an external actuation into propulsive power output can be critical. 
Here we use a microswimmer composed by a self-assembled ferromagnetic rod and a paramagnetic sphere and directly determine its swimming efficiency when it is actuated by a swinging magnetic field. Using fast video recording and numerical simulations we fully characterize the dynamics of the propeller and identify the two independent degrees of freedom which allow its propulsion. We then obtain experimentally the Lighthill's energetic efficiency of the swimmer by
measuring the power consumed during propulsion 
and the energy required to translate the propeller at the same speed.  Finally, we discuss how the efficiency of our microswimmer could be 
increased upon suitable tuning of the different experimental parameters.}}\\


\rmfamily 


The realization of faster and smaller  
micro/nanopropellers  
is an active topic 
with direct applications 
in the emerging fields
of drug delivery~\cite{Kim13,Kat13,Wang14},
microsurgery~\cite{Nel10,Pey13} and
lab-on-a-chip technology~\cite{San11,Wan12}.
The main challenge arises from the 
low Reynolds number regime in which these micropropellers operate, where viscous forces dominate 
over inertial ones and the Stokes equation becomes time
reversible. Thus, in order to attain propulsion, these prototypes should avoid
reciprocal
motion, namely periodic backward and forward displacement.
As described by Purcell, this condition can be satisfied 
by a minimum of two independent degrees of freedom, 
when their change in time describes a finite area 
in the parameter space~\cite{Pur97}.  

Recent years have  witnessed a variety of  theoretical propositions~\cite{Marc13}, and a number of
experimental advances~\cite{Bec16} which have provided 
multiple micropropeller prototypes 
based on different propulsion mechanisms, including
chemical reactions~\cite{Pax04,How07}, magnetic~\cite{Tie08,Zha09,Sne09,Fis11} or light fields~\cite{Vol11,Pal16}. 
Most of these prototypes have been 
characterized and compared mainly 
in terms of their achievable speed, 
a simple observable that can be calculated directly 
from particle tracking.  
However, choosing the correct propulsion scheme for a specific task 
requires a suitable measure of the propeller efficiency
not limited to the sole speed. 
The control of swimming efficiency is of paramount importance both in artificial microswimmers to, e.g. avoid excessive heating due to dissipation, and in microorganisms to optimize energy 
release~\cite{Ost11}.

In $1975$ Lighthill quantified the swimmer efficiency in converting 
input energy to thrust power through a single parameter  $e$, which 
compares the external power $\langle \Phi \rangle$, required to induce a mean velocity $V$ in a medium of viscosity $\eta$, to the power needed to rigidly drag the swimmer at the same speed with an external force $F_{drag}$~\cite{Lig75},
\begin{equation}
e = \frac{F_{drag} V}{\langle \Phi \rangle}
\end{equation}
Although other measures of the efficiency have been proposed~\cite{Sha89, Ost11}, especially to account for collective ciliary motions, 
the Lighthill energetic efficiency remains the standard measure to account for swimming efficiency of single propellers at the microscale.
This parameter has been employed in different
theoretical works to analyze
the performance of simple
artificial designs like three-link flagella~\cite{Bec03,Pas12,Wie16},
three-sphere swimmers~\cite{Ram08},
squirmers~\cite{Ishi14}, necklace-like propellers~\cite{Raz08} and
undulating magnetic systems~\cite{Emi16}.   
However, to date Lighthill's efficiency has been only  experimentally estimated in few biological systems~\cite{Cha06,Gli06}, where  the complex flagellar dynamics precludes the possibility to directly determine the relevant degrees of freedom of the  microswimmer, and indirect methods such as optical tweezers, are used.
 
In this article 
we directly measure the energetic Lighthill efficiency of a minimal magnetic microswimmer cyclically actuated by an external magnetic field. 
The microswimmer is composed by a 
ferromagnetic nanorod 
and a paramagnetic microsphere
which self-assemble due to magnetic dipolar interactions. 
The pair is subjected to a swinging magnetic field
that forces the rod to tilt and to slide at the surface of the microsphere. 
Such an actuating field would not induce propulsion for a single body 
swimmer in a viscous fluid, unless composed by a flexible tail~\cite{Dre05}. 
The simplicity of the microswimmer allows us to show that its locomotion is enabled by only two independent degrees of freedom, 
which are experimentally identified and characterized. 
The precise measurement of the microswimmer dynamics allows for   a direct, accurate and systematic measurement  of its energetic Lighthill efficiency. 
By combining real space/time experiments
with theory and numerical simulations,
we determine speed and efficiency over a large
range of frequencies and amplitudes. We will first describe the microswimmer and indicate how we exploit its simplicity to describe 
its behavior both experimentally and theoretically. We exploit subsequently this detailed control to quantify its efficiency and characterize its regime of motion.

Our hybrid microswimmer is composed by 
a spherical paramagnetic particle 
of radius $R= 1.5 \rm{\mu m}$ 
and a ferromagnetic Nickel nanorod
that is synthesized via template-assisted electrodeposition.
The latter has a length of $L= 3 \rm{\mu m}$, 
a diameter of $D=400 \rm{nm}$
and displays a permanent magnetic moment along its 
long axis, $m_n=3.7 \cdot 10^{-11} \rm{A m^{-2}}$.
In contrast, the spherical particle is characterized by a magnetic volume susceptibility $\chi=0.21$,
and under an external field $\bm{B}$ acquires an induced moment $\bm{m}_p=4\pi R^3 \chi \bm{B}/(3 \mu_0)$
which points along the field direction, being $\mu_0=4 \pi \cdot 10^{-7} \rm{Hm^{-1}}$.
Both elements are dispersed in highly deionized 
water and allowed to sediment above a glass substrate, which is placed at the center of a set of orthogonal coils 
arranged on the stage of a brightfield optical microscope, more details are in the Material and Methods section.
To further understand the experimental data, we perform numerical simulations of a simple theoretical model of the microswimmer which captures the geometry of the particles and considers
their mutual magnetic and hydrodynamic interactions, 
see Materials and Methods section for more details.

\begin{figure}[h]
\begin{center}
\includegraphics[width=\columnwidth,keepaspectratio]{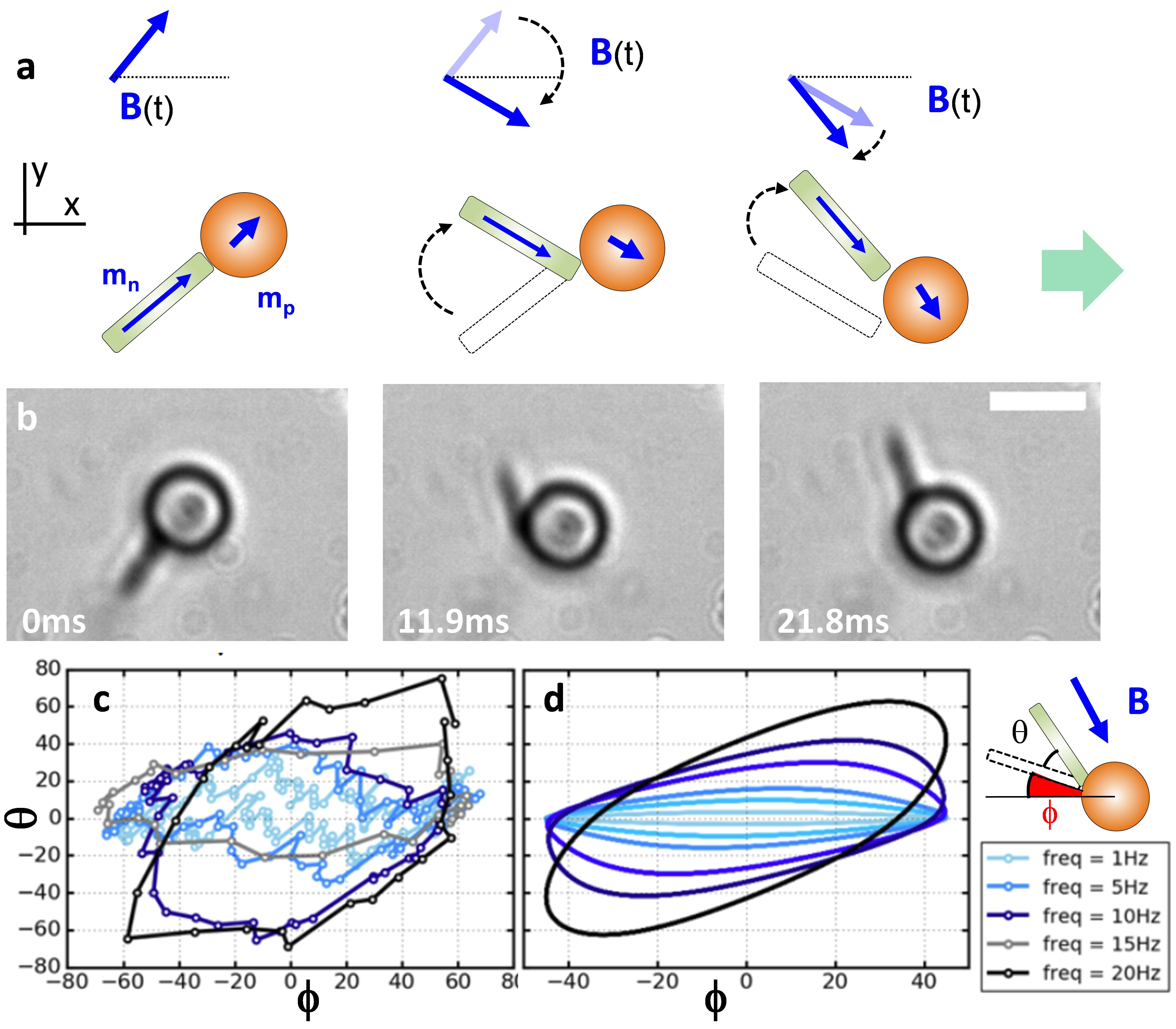}
\caption{(a) Sequence of schematics showing the colloid-nanorod pair subjected to 
the swinging magnetic field $\bm{B}(t)$ during half-cycle,
$m_n$  and $m_p$ denote, respectively, the permanent and the induced moments 
of the nanorod and the colloid. 
(b) Corresponding microscope images
showing relative movement of the pair during half cycle. 
The pair is propelled by a field with frequency $\nu=20 \rm{Hz}$
and amplitudes $B_x=2.15 \rm{mT}$, $B_y= 2.74\rm{mT}$ 
and recorded at $504$ fps, see corresponding VideoS1 and VideoS2 in the Supporting Information.  
The scale bar is $5\rm{\mu m}$.
(c,d) Cycles in the ($\phi,\theta$) plane 
at different frequencies of the applied field for 
experiments (c) and numerical simulation (d).
The small schematic on the right provides the definition of the rotational angle of the nanorod $(\theta)$, and the angle between the $x$-axis and the contact point between the nanorod and the particle, $(\phi)$}.
\label{fig_1}
\end{center}
\end{figure}
We actuate our prototype 
by using a swinging magnetic field composed by 
an oscillating component $B_y$ with frequency $\nu$
and a perpendicular, static field $B_x$, 
$\bm{B}\equiv [B_x, B_y \sin{(2\pi \nu t)}]$.
The effect of the applied field on the relative 
displacement of the two elements during 
a half-cycle, $t\in[1/(2\nu),3/(2\nu)]$, 
is shown in Fig.1(a).
Initially, both the 
permanent, $m_n$, and 
the induced, $m_p$, moments are aligned with the 
field direction, $B_x,B_y>0$. 
As $B_y$ oscillates, $m_p$ follows 
instantaneously the external field. The nanorod rotates  an angle $\theta$ about the contact point with the paramagnetic particle due to the torque exerted by the field.
Simultaneously, the attractive dipolar interaction between the 
pair forces the rod to slide over the particle surface  to align with $m_p$,
minimizing the magnetic energy of the pair. 
Such sliding motion can be characterized by a second degree of freedom $\phi$ (see Fig 1), which represents the angle between the $x$-axis and the contact point between the tip of the rod and the particle.  The uncoupled rotation and sliding 
motions
are able
to break the time-reversal symmetry of the flow, leading to the microswimmer net translation.
This mechanism is observed in our numerical simulations and demonstrated experimentally in Fig.1(b), where the pair moves with the microsphere in front of it
for these geometric parameters. Therefore, 
the propulsion of the microswimmer is allowed by the
two decoupled rotational modes and it  does not require the presence of a bounding wall, as opposed to the mechanism reported in  Ref.~\cite{Jos18}.

We record high frame-rate videos to directly measure the two angular degrees of freedom $(\phi,\theta)$.
The non-reciprocity of the cyclical motion is evident from the trajectories 
in the parameter space $(\phi,\theta)$ shown in  
Fig.1(c) for experiments, and in Fig.1(d) for simulations. Indeed, 
the trajectories follow different paths
in the first and second half-cycles of the actuating field $\mathbf{B}(t)$, 
describing a closed 
asymmetric region. As the frequency of the field is reduced, the 
rotation and sliding motions of the rod become more 
synchronized and the deformation approaches reciprocity. 
As a result, the first and second half-cycles of the 
trajectories shown in Figs.1(c,d) approach one another, 
coming close to a straight line for the lowest frequencies.   

\begin{figure}[ht]
\begin{center}
\includegraphics[width=\columnwidth,keepaspectratio]{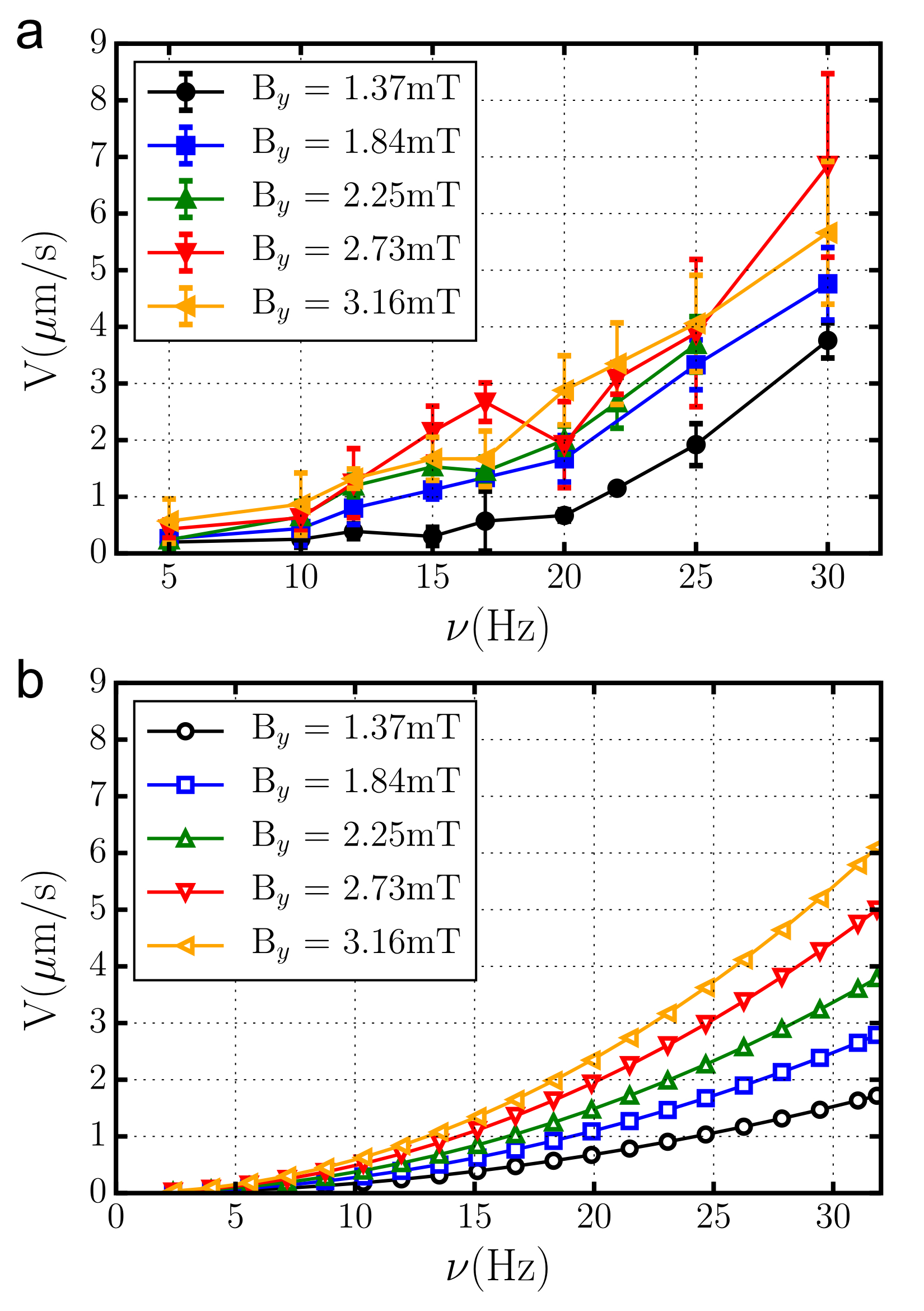}
\caption{(a,b) Velocity of the magnetic propeller 
as a function of the
frequency $\nu$ for different amplitudes 
of the applied field from experiments (a) 
and simulations (b). In both cases the 
static field 
component is kept constant to $B_x= 2.15\rm{mT}$.}
\label{fig_2}
\end{center}
\end{figure}

Fig.2(a) displays the resulting
propulsion speed as a function of frequency and amplitude $B_y$
of the driving magnetic field, with fixed longitudinal component $B_x= 2.15\rm{mT}$. 
The experimental curves follow the same trends obtained 
in numerical simulations of the system shown in Fig.~2(b). Both set of results demonstrate that
the propeller velocity increases with the frequency of the 
transverse oscillating field,
consistent with the observed increase in the area
enclosed by the $(\phi,\theta)$ trajectories.

The microswimmer displacement is determined by the motion of the contact point of the rod and the sphere. This motion can be resolved using resistive force theory~\cite{kim_karrila} and Stokes's law to account for the overdamped interaction of the rod and spherical colloid with the fluid, respectively.  
In this formulation, the propulsion speed is controlled
by three dimensionless parameters, the ratio between the particles sizes
$\delta = R/L$,  $\alpha_{ind} = \mu_0 m_n/(R^3 B_x)$  that compares the magnetic field 
induced by the ferromagnet on the paramagnetic particle with the external field, and the susceptibility $\chi$ of the paramagnetic particle. The propeller dynamics can be solved perturbatively~\cite{Emi14,Emi16,Wie16}
for a weak  oscillating transverse field, $ B_y/B_x \equiv \epsilon \ll 1$. 

The averaged propeller velocity $V$, to leading order in $\epsilon$ reads
\begin{equation}\label{Eq1}
 V/V_0 = \epsilon^2 \frac{ b_1 (\nu/\nu_0)^2}{a_4(\nu/\nu_0)^4 + a_2(\nu/\nu_0)^2 + a_0} + \mathcal{O}(\epsilon^4)  \,,
\end{equation}
where $V_0 = R \nu_0$, and $\nu_0 = \frac{9 B_x m_n}{\pi \eta L^3}$ is a characteristic frequency of the problem. $b_1, a_0, a_2, a_4$ are
coefficients which depend on $\delta, \alpha_{ind}$ and $\chi$ (see Supporting Information). 
Eq.~\ref{Eq1} shows a non-monotonic dependence on frequency, leading to an optimal propulsion at $\nu_{V}^{max}=\nu_0 (a_0/a_4)^{1/4}$. For the experimentally  relevant frequency range, one can approximate
$V/V_0 =  \epsilon^2 (b_1/a_0) (\nu/\nu_0)^2$, exhibiting a quadratic dependence on the frequency
of the transverse field, consistent with the results obtained both experimentally and computationally, see Fig.~\ref{fig_2}.
Although the experiments need to be performed in conditions where $ B_y/B_x \approx 1$ to be able to track in detail the motions of the magnetic particles, the 
leading order solutions of the theoretical model in $\epsilon$, Eq.~\ref{Eq1} (and Eq.~\ref{Eq:eff} below), provide insights on the functional form and relevant parameters of the problem. The perturbative analytical solutions, Eq.~\ref{Eq1}, are validated by comparing with numerical simulations of the model, exhibiting good agreement even for $ B_y/B_x \approx 1$. 

In order to extract the Lighthill efficiency, $e$, from the experimental data, 
we measured independently the external power $P_B$ exerted on the system, and the 
equivalent power $P$ required to translate the propeller. 
Since the only work exerted by the external field is due to the torque on the ferromagnetic rod (which is then partly transferred to the sliding motion), we can express the instantaneous external power as $P_B(t) =  \bm{\tau}_B(t)\cdot \dot{\bm{\alpha}}$, 
being $\bm{\tau}_B(t) = {\bf m}_n(t)\times{\bf B} (t)$ the instantaneous torque applied on the rod, and $\dot{\alpha}$ the rod angular velocity. 
The average power in one period of the magnetic field is given by $\bar{P}_B = \nu\int_0^{1/\nu} P_B(t) dt$. 
This quantity is determined from
the orientation and angular velocity of the rod at each instant of time, see Materials and Methods for more details. We calculate the phase between ${\bf m}_n$ and ${\bf B}$  by coupling to a light emitting diode the signal of the ac current flowing through the Helmholtz coils. 
The power needed to rigidly drag the swimmer at a given speed $v$ is obtained by using controlled magnetic gradients, $P =v {\bf m}_n \cdot \nabla{\bf B}$.
The resulting experimental data are shown in Fig.3(a), where $e$ is measured in the frequency range $\nu \in [10,30] \rm{Hz}$ and at different 
amplitudes $B_y$ with fixed $B_x=2.15 \rm{mT}$.
In the efficiency calculation we neglect the contribution due to the rotational motion of the paramagnetic sphere. From the analysis of the experimental videos we find that that such rotation was $\sim 5^{\circ}$ and its contribution negligible respect to the rotation of the rod.

\begin{figure}[ht]
\begin{center}
\includegraphics[width=\columnwidth,keepaspectratio]{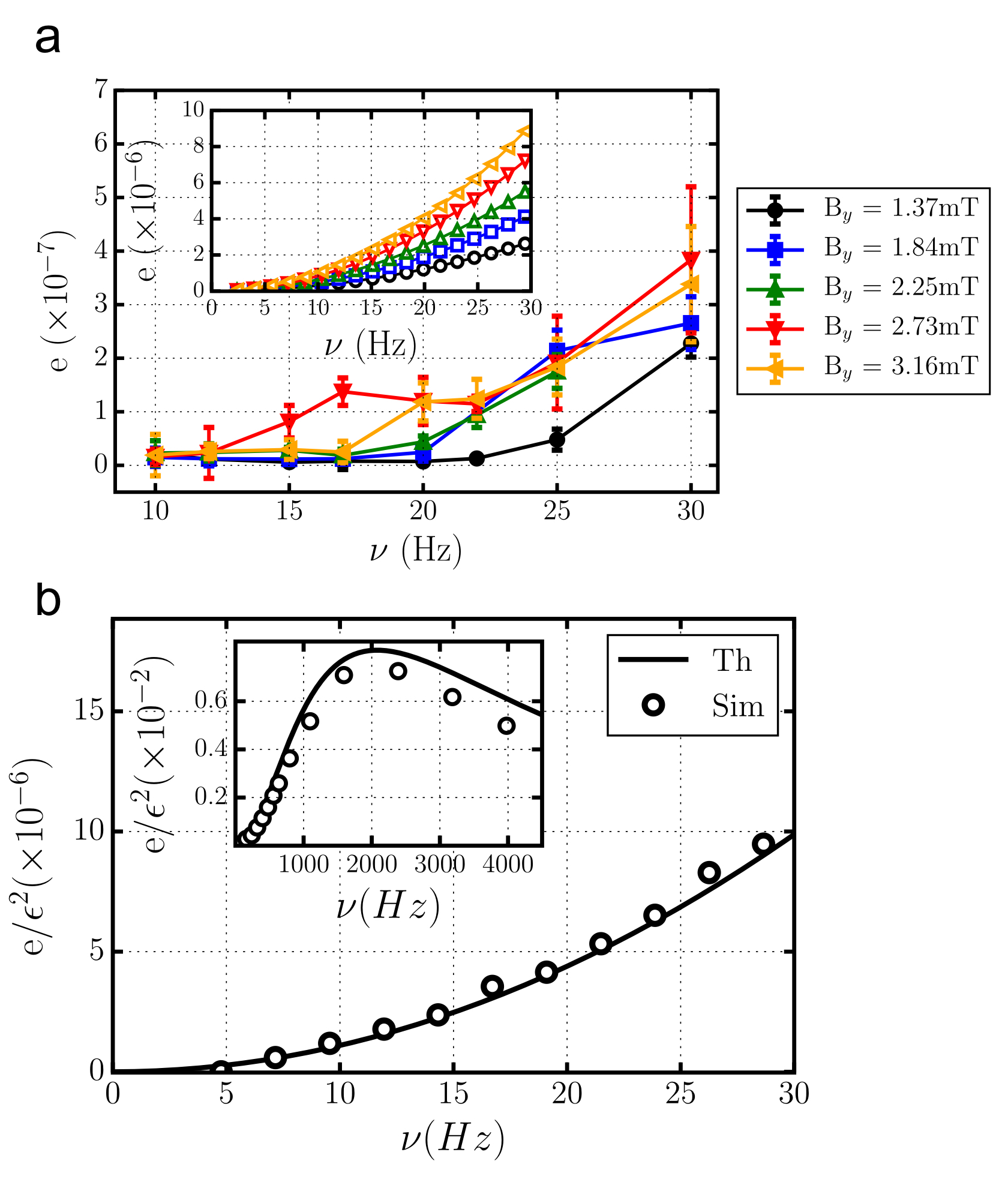}
\caption{(a) Propeller efficiency $e$  
versus frequency $\nu$ of the swinging field 
for different amplitudes $B_y$ as measured in the experiments. Inset shows results for the efficiency obtained from simulations. (b) Normalized efficiency $e/\epsilon^2$ versus frequency $\nu$ as obtained from the theoretical model (solid line) and simulations (dots). 
Inset shows $e/\epsilon^2$ for a wide range of frequencies to visualize a peak at $\nu_{e}^{max} = 1776\rm{Hz}$.}
\label{fig_3}
\end{center}
\end{figure}

The perturbative solution of the dynamical model provides an analytical expression for the Lighthill efficiency at small driving amplitudes $\epsilon$,
\begin{equation}\label{Eq:eff}
e = \epsilon^2 \frac{ n_2 (\nu/\nu_0)^2}{d_6(\nu/\nu_0)^6 + d_4(\nu/\nu_0)^4 + d_2(\nu/\nu_0)^2 + d_0} + \mathcal{O}(\epsilon^4)\,,
\end{equation}
which reduces to $e = \epsilon^2 (n_2/d_0) (\nu/\nu_0)^2$ in the range of experimentally accessible frequencies. The coefficients $n_2$, $d_6$, $d_4$ and $d_2$ are given by the parameters of the problem  $\delta, \alpha_{ind}, \chi$. 
Figs.~\ref{fig_3} show a qualitative agreement between the measured values of the Lighthill efficiency and the predictions from simulations and the dynamical model.

We note that the quantitative discrepancies between the experiments and the computational predictions can be attributed to different factors. Specifically,  paramagnetic colloids, composed of magnetic domains in a non-magnetic matrix, can present a residual magnetic anisotropy \cite{Jan09}. As a result, the rotation of the spherical colloid due to the interaction with the external field and the field induced by the ferromagnetic rod can contribute to the  dissipation, without affecting the overall swimmer propulsion. In addition, as a consequence of such interaction, the ferromagnetic rod could acquire a higher dephasing with the external field, resulting in a larger input power. Other factors such as the interaction with the surface and the action of gravity on the couple could also be a source of discrepancy.

Eq.~\ref{Eq:eff} predicts the existence of an optimal efficiency at a frequency $\nu_{e}^{max}$ ($\sim1776$~Hz for the parameters of the experimental system), where the efficiency is over three orders of magnitude higher than 
the one measured at the experimentally accessible frequencies (see inset of Fig.~\ref{fig_3}).  
Additionally, the analytical model allows for an optimal design of the microswimmer that  maximizes 
its efficiency for a given field $\mathbf{B}(t)$.
While the efficiency is not very sensitive to $\delta$ and $\alpha_{ind}$, it exhibits a strong dependence on the susceptibility of the paramagnetic
particle (see Supporting Information). Indeed, Eq.~\ref{Eq:eff} predicts that the efficiency of the microswimmer has a maximum for a susceptibility $\chi_{max} \approx 0.32$ if one keeps the other parameters of the microswimmer constant,
for which the efficiency is more than an order of magnitude higher.
Note that for $\chi \gg 1$, the phase difference between the rotation and sliding  motions of the ferromagnetic rod vanishes generating a reciprocal motion. On the other hand, for $\chi \ll 1$ the sliding motion does not occur, leaving only one independent degree of freedom and thus no propulsion.
Note also that the frequency of maximum efficiency is inversely proportional to the viscosity of the solvent, which provides a possible route to optimize the efficiency of the swimmer.

In conclusion, we have determined the Lighthill energetic efficiency of a  minimal  self-assembled microswimmer 
composed by a paramagnetic microsphere and a ferromagnetic nanorod,  and  have developed a theoretical scheme  that captures the essential dynamics of the system. 
The simplicity of the synthetic microswimmer allows a quantitative and precise measurement of its efficiency, showing a value smaller than flexible magnetic swimmers~\cite{Gauge2006} but larger than other artificial non-propelling engines~\cite{Sch18}. We show that the efficiency of the microswimmer is much more sensitive to  the internal degrees of freedom, which are not relevant to determine  its propulsion speed. 
The theoretical study has shown the existence of an optimal efficiency, which is three orders of magnitude larger than the  one obtained experimentally.  
While the peak efficiency could be observed only in the model given the experimental limitations, 
we show that it could be controlled by varying the field parameters or the magnetic susceptibility
of the paramagnetic colloid. 
The realization of 
minimal, artificial prototypes that avoid the complexity of 
biological systems and that can be controlled through their 
independent degrees of freedom, is still 
an open challenge in the field,
in spite of the amount of theoretical propositions.  
More in general, when compared to other prototypes driven by external fields, 
magnetic microswimmers
have the advantage of being easily controlled and steered trough the fluid 
without altering the composition of the dispersing medium, all features that make them rather appealing for practical applications. 
Finally, the possibility to characterize the swimming efficiency of
nano/micro propellers in terms of the constituent degrees 
of freedom represent a key issue
in many technological contexts~\cite{Ghosh2013}.

\section*{Conflicts of interest}
There are no conflicts to declare'.

\section*{Acknowledgements}
This work has received funding from the
Horizon 2020 research and innovation programme,
Grant Agreement No. 665440.
F.S. acknowledges support from MINECO under
project FIS2016-C2-1-P AEI/FEDER-EU.
I.P. acknowledges support from MINECO under
project FIS2015-67837-P and Generalitat de Catalunya under project 2017SGR-884 and SNF Project No.
200021-175719.
P.T. acknowledges the 
European Research Council (ENFORCE, No. 811234), MINECO 
(FIS2016-78507-C2-2-P, ERC2018-092827) 
and Generalitat de Catalunya under program "Icrea Academia".

\section*{Materials and Methods}

\subsection*{Experimental System and Methods}

The Ni nanorods are synthesized by template-assisted
electrodeposition from a single electrolyte, $0.5 \rm{mol \cdot dm^{-3}}$
NiCl$_2$ solution (Sigma Aldrich), prepared with distilled
water treated with a Millipore (Milli Q system). The
electrosynthesis was conducted potentiostatically using a
microcomputer-controlled potentiostat/galvanostat Autolab
with PGSTAT30 equipment, GPES software and a three electrode
system. A polycarbonate (PC) membrane with pore
diameter $\sim 400 \rm{nm}$ (Merck-MilliPore) and sputter-coated
with a gold layer on one side to make it conductive is used as
the working electrode. The reference and the counter
electrodes are a Ag/AgCl/KCl ($3\rm{mol \cdot dm^{-3}}$) electrode and a
platinum sheet respectively. After synthesis, the Ni nanorods
are released from the membrane by first removing the gold
layer with a I$_2$/I$^-$ aqueous solution, and then by wet etching of the PC membrane in CHCl$_3$. Nanorods are then subsequently
washed with chloroform ($10$ times), chloroform-ethanol
mixtures ($3$ times), ethanol ($2$ times) and deionised water 
($5$ times). Finally, sodium dodecyl sulphate (Sigma Aldrich) is
added to disperse nanorods. The typical length of the fabricated Ni
nanorods used in this study is around $L =  3 \rm{\mu m}$. Structural
and morphological analysis were carried out with scanning
and high-resolution transmission electron microscopes. The
permanent moment of the nanorod is measured by following
its orientation under a static magnetic field, as described in
previous works~\cite{Mar16,Mart16}.
The value obtained for the magnetic moment of the ferromagnetic
rod is $m_n = 3.7\times10^{-11} \rm{Am^{-2}}$. 

The spherical colloids used are paramagnetic microspheres with radius $R = 1.5 \rm{\mu m}$, $\sim 15 \%$ iron oxide content and surface carboxylic groups (ProMag PMC3N, Bang Laboratories). 
The particles are characterized by a magnetic volume susceptibility equal to $\chi=0.21$, as measured in separate experiments~\cite{Jos18}. 
The particles and the nanorods are dispersed in highly deionized water (MilliQ, Millipore) and allowed to sediment above a glass substrate. 
The substrate is placed in the center of five orthogonal coils arranged on the stage of a light microscope (Eclipse Ni, Nikon), equipped with a Nikon $100\times$ objective with $1.3$NA. The coils are connected to a wave generator (TGA1244, TTi) feeding a power amplifier (IMG AMP-1800). 
The particle dynamics are recorded with a CCD camera (scA640-74fc, Basler) working at around $75$ frames per second (fps), with a CMOS camera (MQ003MG-CM, Ximea) working at $500$ fps, or in color at $325$ fps (acA640-750uc, Basler). 

\subsection*{Measurement of the phase of the field.}

To measure the phase between the instantaneous value of the applied field and the orientation of the propeller,
we modify the experimental set-up by introducing two LEDs to the optical path, just above the observation objective. The two LEDs are connected in an anti-parallel configuration to an alternate current (AC) voltage source, which is produced by the same waveform generator that powers the magnetic coil system. We use a phase lock program to synchronize the oscillations coming from the two signals. In this configuration, the green LED emits light during the positive cycle
of the applied field, while the red LED emits on the negative one. The tube lens of the objective allows to distribute the colored light over the whole sample view. Even if the transmitted intensity appears as relatively small, it can be distinguished from the experimental image. From the color video in RGB format, we calculate the average value of all the pixels in the red and in the green channels as a function of time. 

We then perform a least squares fit using the function 
$f(t) = A + \frac{B}{2} \left(\sin(2\pi ft + \phi)+
\left|\sin(2\pi ft + \phi)\right|\right)$ from which we extract the phase $\phi$, being $A$ and $B$ two amplitudes. The value of $\phi$ allows us to calculate an instantaneous value of the field for each frame. 
Further, we  track three points of the swimmer using the public program  ImageJ (National Institutes of Health). 
These points are the outermost tip of the nanorod, the point of contact between the nanorod and the colloidal particle, and the center of the colloidal particle. From these three points, we extract the relative angles, and using the instantaneous direction  of the applied field $\bf B$, we have all the information 
over the different degrees of freedom involved.

\subsection*{Numerical simulations}

The paramagnetic particle is modeled as a spherical bead of radius $R$ with an induced magnetic 
moment ${\bm m}_{p}$ located in its center. 
The ferromagnetic rod is described as a group of $N$ equally spaced spherical beads of diameter $D$ along a straight line. Every bead carries a 
fixed magnetic moment ${\bm m}_{n}/N$ directed along the axis of the rod. The external magnetic field exerts a torque on the ferromagnet
${\bm \tau}_B =  {\bm m}_n \times {\bm B}$, which is implemented as an artificial force pair applied perpendicular to the axis of the rod. The paramagnetic particle and the ferromagnetic rod interact through magnetic dipolar interactions, which are calculated as a sum of dipolar interactions between the paramagnetic sphere and the $N$ beads composing the ferromagnetic rod. 
In addition, the spherical beads composing the swimmer interact with other beads and with the bounding plane at $z=0$ through short ranged steric interactions, which are described using the Weeks-Chandler-Andersen (WCA) potential \cite{WCA}. Such interactions provide the beads with extension and prevent overlaps.

Due to the dimensions of the swimmer we assume that its dynamics is overdamped and governed by the hydrodynamic drag with the viscous fluid. 
The interaction of bead $i$ of the microswimmer with the viscous fluid is given by the hydrodynamic friction
force 
\begin{equation}
 {\bf F}_{H,i} = -\gamma_i({\bf v}_i - {\bf u}({\bf r}_i))\,,
\end{equation}
where $\gamma_i$ is the bead's friction coefficient, ${\bf v}_i$ its velocity, and ${\bf u}({\bf r}_i)$ 
is the induced fluid flow at the bead's position. The flow field ${\bf u}({\bf r})$ is generated by the action of the net (non hydrodynamic) forces 
on the different elements of the microswimmer, which is treated in the far field approximation. In fact, we approximate the hydrodynamic behavior of the particles composing the microswimmer
to that of point particles, which gives
\begin{equation}
 {\bf u}({\bf r}_i) = \frac{1}{8\pi \eta}\sum_{j} G({\bf r}_i;{\bf r}_j) \cdot {\bf F}_j\,.
\end{equation}
Here, ${\bf F}_j$ is the non-hydrodynamic force acting on particle $j$, $\eta$ is the viscosity of the fluid,  
and $G({\bf r}_i;{\bf r}_j)$ is the Green's function of the Stokes equation. 
We have considered both the case of an unbound fluid, for which we use the Oseen tensor $G^{Oseen}({\bf r}_i;{\bf r}_j)$, 
and also the case with a no-slip, flat and infinite boundary for the fluid flow, for which we use the Blake tensor \cite{Bla71} $G^{Blake}({\bf r}_i;{\bf r}_j)$.
The actuation on our microswimmer by the applied external fields dominates over thermal effects (${\bf m}_n B \gg k_B T$). For this reason, our numerical simulations do not consider the effect of temperature on the dynamics of the magnetic couple.

The dynamics of the swimmer evolves following Newton's equations of motion, which are solved using a Verlet
algorithm adapted for cases with forces which depend on the velocity \cite{Bai09}. 
\\

\scriptsize{
\bibliography{bibliography} 
\bibliographystyle{rsc} } 

\end{document}